\documentstyle[multicol,psfig,aps,prl]{revtex}

\begin{document}

\draft
\title{\bf{ Vicinal Surfaces, Fractional Statistics and Universality}}
\author{ Somendra M. Bhattacharjee\cite{eml1}} 
\address{Institute of Physics, Bhubaneswar 751 005, India}
\author{Sutapa Mukherji\cite{eml2}} 
\address{ Department of Physics and Meteorology, Indian Institute of
  Technology, Kharagpur 721 302,India}
\onecolumn
\date{\today}
\maketitle
\widetext
\begin{abstract}
  We propose that the phases of all vicinal surfaces can be
  characterized by four fixed lines, in the renormalization group
  sense, in a three-dimensional space of coupling constants. The
  observed configurations of several Si surfaces are consistent with
  this picture.  One of these fixed lines also describes
  one-dimensional quantum particles with fractional exclusion
  statistics.  The featureless steps of a vicinal surface can
  therefore be thought of as a realization of fractional-statistics
  particles, possibly with additional short-range interactions.
\end{abstract}
\begin{multicols}{2}
  
A crystal surface cut at a small angle to a symmetry direction is
called a vicinal surface\cite{cayley}. Such a miscut surface consists
of terraces of the symmetry plane, separated by monatomic steps
running across the sample in a preferred direction dictated by the
cut. The density of the steps (number of steps per unit transverse
length), $\rho$, is related to the miscut angle $\theta$ $(\rho \sim
\tan \theta)$. In thermal equilibrium, the configuration of the
surface is determined by the steps and their interactions.  Vicinal
surfaces constitute a special class of objects in low dimensional
statistical mechanics and are useful as substrates in many
technological and experimental situations. An understanding of the
phases and phase transitions of vicinal surfaces is therefore
important.
  
Various types of behaviors of vicinal surfaces are known. For example,
it is known that vicinal Si(111) surfaces have single steps or
triplets \cite{will93} but never pairs, while pairs of steps are seen
on Si(001) \cite{webb}.  An unusual phenomenon occurs in the case of
Si(113) where a uniform-step-density phase, on cooling,
phase-separates into a flat surface and a phase with a large angle
(i.e., a high density of steps) \cite{song94,sudoh98}. In other words,
a small-angle vicinal Si(113) surface becomes thermodynamically
unstable.  The coexistence curve, on the
temperature-versus-miscut-angle orientational phase diagram, ends at a
tricritical point with the shape\cite{song94} given by an exponent
$\beta = 0.42$-$0.54$.  A variant of the phase diagram has also been
reported in Ref. [4b]. Is there a natural way of characterizing this
zoo of vicinal surfaces?
  
In this paper our aim is to develop a long-distance, universal
behavior for vicinal surfaces, where microscopic details like the
specific material, the lattice structure, surface reconstruction if
any, etc., do not play a direct role. This is best done in a continuum
approach\cite{jaya,smb96,lass96}.  The steps are treated as
fluctuating elastic strings (directed polymers), and all the effects
of the surface go into the elastic energies of the steps and their
effective interactions.  Therefore, our approach is valid for any
vicinal surface that can be characterized by featureless, wandering
steps.

A major interaction of the steps is known from theory
\cite{forceth,jaya} and experiments \cite{forceexp} to be $\mid {\bf
  r}_i(z)-{\bf r}_j(z)\mid^{-2}$ where $z$ is the special direction of
the steps and ${\bf r}_i$ is the transverse $d$-dimensional coordinate
of a point at $z$ along the length of the $i$th step.  (Our interest
is, of course, at $d=1$.)  This long-range interaction is generally
repulsive, originating from the elastic effects of the terraces,
although, for metals, dipoles or quenched impurities on the steps can
produce an attractive $r^{-2}$ potential\cite{jaya}. A theoretical
explanation \cite{smb96} of the observed tricriticality\cite{song94}
requires an additional attractive short-range interaction. The
continuum description of Ref. \cite{smb96} is justified a posteriori
by the existence of a renormalization group fixed point with a
diverging length-scale (see also Refs. \cite{lass96,shenoy98}).

In an effective Hamiltonian for the steps, the interaction need not be
restricted to just pairwise-additive potentials, and, in a
renormalization-group (RG) approach, one should include allowed
marginal operators.  We keep the three-body short-range interaction
since it is marginal at $d = 1$ (see below) \cite{smbjj,smb96}. All
$m$-step interactions are irrelevant for $m > 3$ at $d = 1$. The
Hamiltonian is
\begin{eqnarray}
  \label{eq:1}
H&=& \sum_i   \int_0^N \! \!\!dz \,  \frac{\kappa_i}{2} \dot{\bf
  r}_i^2(z) + 
\sum_{i>j} \int_0^N \!\!\!\! dz \Big [ v_2 \delta_{\Lambda}({\bf
  r}_{ij}(z)) + \frac{g}{ r_{ij}^2(z)} \Big ]\nonumber\\ 
&&
 +v_3  \sum_{i>j>k} \int_0^N\! dz \delta_{\Lambda}({\bf r}_{ij}(z))
\delta_{\Lambda}({\bf r}_{jk}(z)). 
\end{eqnarray}
where $\dot{\bf r}_i(z)={\partial {\bf r}_i}/{\partial z}$, $\kappa_i$
is the elastic constant of the $i^{{\rm{th}}}$ step, each of length
$N(\rightarrow\infty)$, and $g,\ v_2,\ v_3$ are respectively the
long-range two-body, the short-range two-body and the short-range
three-body coupling constants.  The short-range interactions are taken
as contact interactions, and power counting shows that this is
sufficient \cite{comm1}. There is a short-distance cut-off
(reminiscent of the lattice) so that, in Fourier modes,
$\delta_{\Lambda}({\bf q})=1,$ for $\mid q\mid < \Lambda,$ and $0,$
otherwise. For the tricritical point, one takes \cite{smb96,lass96}
$v_2=v_{20}\, (T-T_t)$ but, for generality, we consider both positive
and negative values of all the three parameters $(g, v_2, v_3)$.  The
partition function is then given by the summation of the Boltzmann
factor over all possible configurations with free end-points. Two
special cases of Eq. (\ref{eq:1}) $(v_3=0, \ \rm{and}\ g = 0)$ were
considered in Ref. \cite{smb96} in connection with the tricritical
point seen in Si(113). The physical picture used is that of phase
separating polymers with the critical point of the coexistence curve
coinciding with the binding-unbinding critical point of two steps
(zero density). It was shown in Ref. \cite{smb96,lass96} that $\beta =
0.5$ occurs for $g = 3/4$.

The resemblance of Eq. (\ref{eq:1}) with the Calogero-Sutherland
model\cite{suth,kol} is apparent and we shall discuss some issues
related to this.  The Hamiltonian $(\hbar=1)$ describes a set of
interacting quantum particles which can be chosen as bosons. The
thermodynamic properties of the quantum system can be obtained from
Eq. (\ref{eq:1}) with $N$ as the inverse temperature.  In the limit of
infinite lengths of the steps, the properties of the vicinal surface
are given by the ground state of the corresponding quantum
problem\cite{jaya}.

The main contents of the paper are the following: (a) We show that
for, the $v_3=0$ case, at $g=3/4$ the continuous transition goes over
to a first-order transition\cite{kol,lipow}.  This requires a study of
reunion exponents\cite{fisher84,suta} for the steps with the
long-range interaction.  The change in the order of the transition is
then connected to the observed behavior of Si(113).  (b) We obtain a
fixed-point (or, rather, a fixed-line) description for the above
Hamiltonian, and, from the nature of the fixed points, we argue that
the observed features of various Si vicinal surfaces are described by
these fixed points.  We then conjecture that {\it all vicinal surfaces
  are described by these fixed lines (and the flows) in a
  three-dimensional parameter space}.  (c) It is quite common to use a
quantum description where the steps are treated as fermions, but we
conclude, from the equivalence with the Calogero-Sutherland model on
the $v_3=0$ fixed line, that the steps should rather be treated as
{\it one-dimensional fractional-exclusion-statistics particles with
  only short-range interactions}.

So far as the RG flows of the various parameters are concerned, one
may just restrict oneself to two or three steps (``vertex
functions'').  For a many-step system, one also needs the chemical
potential $\mu$, but its flow is determined by dimensional
analysis\cite{smbjj,kol,nelson}.  In the momentum-shell
renormalization-group approach, short-wavelength fluctuations are
integrated out in a thin shell $(\Lambda/b, \Lambda)$ in Fourier
space, and the effect is absorbed by redefining the parameters.  The
system is then rescaled to its original state, whereby the cutoff goes
back to $\Lambda$. (We choose $\Lambda=1$.)  The change of the
parameters with distance in the long-distance limit is then recovered
from the RG flow-equations.  The spatial rescaling factor is
anisotropic, $r \rightarrow b r$, $z \rightarrow b^{\zeta} z$, and the
choice $\zeta=2$ keeps the elastic constant invariant\cite{nelson}.
For simplicity, we choose $\kappa_i =1,$ for all $i$.

Since the RG transformation is analytic, the long-range term does not
get renormalized.  That there is no renormalization of $g$ turns out
to be rather natural when we discuss the quantum problem.  We just
quote the well-known recursion relations in terms of the dimensionless
variable\cite{kol} $ u_2 = L^{2-d} K_d v_2 + g/(d-2),$ where
$K_d=2\pi^{d/2}/[\Gamma(d/2) (2\pi)^d]$ and $L$ is an arbitrary
length-scale in the transverse direction $(b=1+\delta L/L)$.
Restricting ourselves to $d=1$, the flow equations
are\cite{kol,nelson,smbjj}
\begin{equation}
   \label{eq:5}
   L\frac{ du_2}{dL} = u_2 - u_2^2 + g, \  \  L\frac{dg}{dL} =0,\ \
    {\rm and}  \  \  L\frac{d\mu}{dL} = 2 \mu.
\end{equation}
The flow equation for $u_2$ has two fixed points $u_{\rm{s,u}}^*=\{ 1
\pm[1 + 4g]^{1/2}\}/2$, s,u denoting respectively the stable and
unstable fixed points (Fig. 1a).  The unstable fixed point describes
the binding-unbinding transition of two steps.  The critical point has
a length-scale exponent $\nu_{\perp} = 1/\sqrt{1+4g}$. For a many-step
system this unstable fixed point corresponds to the tricritical point
mentioned earlier, provided the fixed point describes a critical
behavior.  Since the density $\rho$ vanishes at this tricritical
point, the assumption of a single length scale then tells us that
$\rho^{-1}$ should diverge as the length scale.  Therefore, $\beta=
\nu_{\perp} $.  For $\beta = 1/2$, as for Si(113), one would require
$g=3/4$ as obtained in Ref. \cite{smb96}.
 
We now show that $g=3/4$ is indeed a special point.  Let us consider
the free energy \cite{fisher84} of two steps. Configurations of the
steps are of the type shown in Fig. 2. The bubble-like contributions
(region B in Fig. 2a) represent the steps in the high temperature
phase described by the stable fixed point.  It is shown in Ref.
\cite{fisher84} (we, therefore, skip the details) that the nature of
the singularity of the free energy is determined by the decay of the
partition function of the bubble of length $N$, $Z_R (N)\sim
N^{-\psi}$, where $\psi$ is the reunion exponent\cite{suta}.  If
$1<\psi\leq 2$, then the transition is continuous, while for $2<\psi$
it is first order, but with a weak singularity and a diverging length
scale\cite{comm2}.
 
The reunion exponent $\psi$ determines the behavior of the partition
function of two steps starting at the origin and reuniting anywhere in
space\cite{fisher84,suta}.  We need the exponent in the
high-temperature phase characterized by the stable fixed point.  The
zeroth- and the first-order diagrams for the reunion partition
function $Z_R$ are shown in Figs. \ref{fig:2}b,c.  These contributions
require renormalization of the partition function itself, over and
above the renormalization of the parameters.  This extra
renormalization yields a nontrivial $\psi$ different from the Gaussian
value $\psi=d/2$ in $d$ dimensions.  In the limit
$N\rightarrow\infty$, the loop contributions lead to the recursion
relation for $Z_R(N)$ as $L \ {d Z_R}/{d L}= -2(d/2 + u_2) Z_R,$ so
that at $d=1$ at the stable fixed point $u_2^*=u_{\rm s}$, we get
$\psi = 1 + ( 1+ 4g)^{1/2}/2$.  This is an exact result from which we
recover the vicious walker exponent\cite{fisher84} $\psi=3/2$ for
$g=0$. (The details and generalizations to arbitrary number of steps
will be discussed elsewhere.)  We see that $\psi=2$ for $g=3/4$.  This
then establishes \cite{fisher84,lipow} that the transition is first
order for $g>3/4$.
 
In the case of a first-order two-step phase-transition, this may not
be the critical point of the phase-coexistence line of a many-step
system.  In such a case, a phase diagram of the type shown in Fig.
\ref{fig:1}b.2 is plausible, where a first-order line continues from
the zero-density transition point, ending on the coexistence curve
(hidden region in Fig. 1b.2, see figure caption).  The first-order
line represents a tranition from a uniform phase of single steps to a
uniform phase of pairs, and, at the tranition, there will be diverging
fluctuations in the mean spacing between the members of the pairs (for
$3>\psi>2$).  In recent experiments on Si(113) such a phase diagram
has been observed\cite{song94} for certain azimuthal angles of the
miscut direction, though the low-density line is yet to be detected.
 
Let us now consider the effect of the three-body
interaction\cite{smbjj,comm1,lass94}.  Even though $v_3$ is marginal
(dimensionless at $d=1$), there are contributions in $v_3$
renormalization, from both $v_2$ and $g$ (i.e.  $u_2$) via the reunion
exponent of three steps. Note that $v_3$ does not affect $u_2$ or $g$.
Taking into account the first nontrivial contribution (Figs. 2d, e),
the recursion relation for $u_3=L^{1-d} v_3$ is ($d=1$)
\begin{equation}
  \label{eq:7}
 L\frac{d u_3}{d L}= -3 u_2 u_3 - A u_3^2,  
\end{equation}
where the numerical value of the constant $A>0$ is not crucial.  The
fixed points (Fig. 1c) are $u_3^*=0$ and $u_3^*=-3u_2/A$. The
stability of the $u_3^*=0$ fixed-point depends on the sign of $u_2^*$.
At $u_2^*=u_s^*$ (fixed point A), a small $u_3$ is irrelevant, but the
new fixed-point C, $u_3^*= -3u_s/A <0$, is unstable.  In contrast, at
$u_2^*=u_u^*$ (fixed point B), with $g>0$, $u_3$ is relevant, and
$u_3^* =(3|u_u|)/A>0$ is stable.  The flows for any $g>0$ are shown in
Fig. \ref{fig:1}c.

Figure \ref{fig:1}c shows that for certain combinations of $g>0,\ 
u_2$, and $u_3$, the steps in the high-temperature phase behave like
the $u_3=0$ case (region ${\rm P}_{\rm f}$) but for sufficiently large
$u_3<0$ there will be triplets of steps but never pairs (Phase ${\rm
  P}_{\rm 3}$).  The transition will be first-order but with
singularities determined by the $g$-dependent fixed point C of Fig.
1c.  We like to associate Si(111) vicinal surfaces\cite{will93} with
this fixed point and the associated flow, because of similar behavior.
For fixed points B and D, there could be pairings of steps, induced by
a variation in $u_2$. Such pairings are seen\cite{webb}, e.g., in
Si(001). A many-step system will show a tricritical behavior at these
fixed points, as e.g. in Si(113). These two fixed points B and D of
Figs. 1c,d have the same thermodynamic exponents but they differ in
the three-step correlation functions. Moreover, if $u_2$ is a
temperature-like axis, then the temperature deviation is not the
proper scaling variable at D.  Taking this as a reason for the large
range\cite{song94} in the observed exponent $\beta$, we may associate
Si(113) with the fixed point D with $g$ around $3/4$ and Si(001) with
fixed point B.  There is, in fact, no requirement that the long-range
interaction should be exactly at $g=3/4$ for Si(113).  Unfortunately
no systematic studies are available for the dependence of $g$ on the
various miscut parameters.  In case $g$ depends on the azimuthal angle
even slightly, it is possible that for certain azimuthal angles
$g>3/4$ and for these cases, one would see a phase diagram of type
Fig.  1b.2, while, for cases with $g<3/4$, one would see type Fig.
1b.1.

We now make a conjecture that all vicinal surfaces are described by
the RG flows and the fixed lines we obtained in the three-dimensional
parameter space.  Other higher-order short-range interactions are
irrelevant. All universal properties are determined by the fixed
lines, and it is the long-range parameter $g$ that determines the
universality class.

Let us now go back to Eq. (\ref{eq:1}) as a quantum-mechanical
problem.  With $v_3=0$, the connection of this Hamiltonian with the
Calogero-Sutherland model\cite{suth} has been noted in the
past\cite{kol,lass96}.  The quantum problem is described by the stable
fixed line and the negative-$g$-part of the unstable line in Fig. 1a.
Furthermore, the Calogero-Sutherland model describes\cite{murthy} a
gas of noninteracting particles obeying Haldane-Wu fractional
exclusion statistics\cite{haldane}.  The long-range $1/r^2$ potential
is best thought of as a statistical interaction, with $g$ determining
the statistics of the particles.  We have established this directly in
our approach by a computation of the quantum second virial coefficient
along the fixed lines of Fig. 1a, since the second virial coefficient
was shown, from general considerations, by Murthy and
Shankar\cite{murthy} to determine the statistics uniquely. We wish to
discuss this technical issue elsewhere.  Suffice it to say here that
the non-renormalization of $g$ in our RG approach is a direct
manifestation of the fact that the statistics of the particles is
independent of length scale and {\it hence a renormalization-group
  invariant}.  This correspondence therefore tells us that the steps
on vicinal surfaces are an analog realization of one-dimensional
fractional-statistics particles.  Though fermions are extensively used
to study the equilibrium properties of steps, the latter are better
represented by fractional-statistics particles with only short-range
interactions.

Our main results have already been summarized. It would be interesting
to study the case of unequal elastic constants of the steps, and the
regions not describable by any fixed point, where other details of the
surface, like lattice periodicity, may be important\cite{shenoy98}.
We end with the suggestions that ({\it i}) attempts be made to
determine accurately the long-range (or statistics) parameter $g$ and
its dependence on the miscut features, as e.g. azimuthal angle, for
various surfaces and materials, ({\it ii}) measurements be done for
three-step correlations, and ({\it iii}) the thermodynamic behavior of
a finite number of steps be studied even if in finite geometries (by
putting barriers).  Most interesting would be the special case of
$g=-1/4$ which, in the surface context, is the point of coalescence of
the stable uniform-density-phase fixed point and the unstable critical
point, and, in the quantum context, is the semion, exactly half way
between boson and fermion.  Can one find such a surface of a metal?

\end{multicols}                                                     
\newpage
\vbox{
\psfig{file=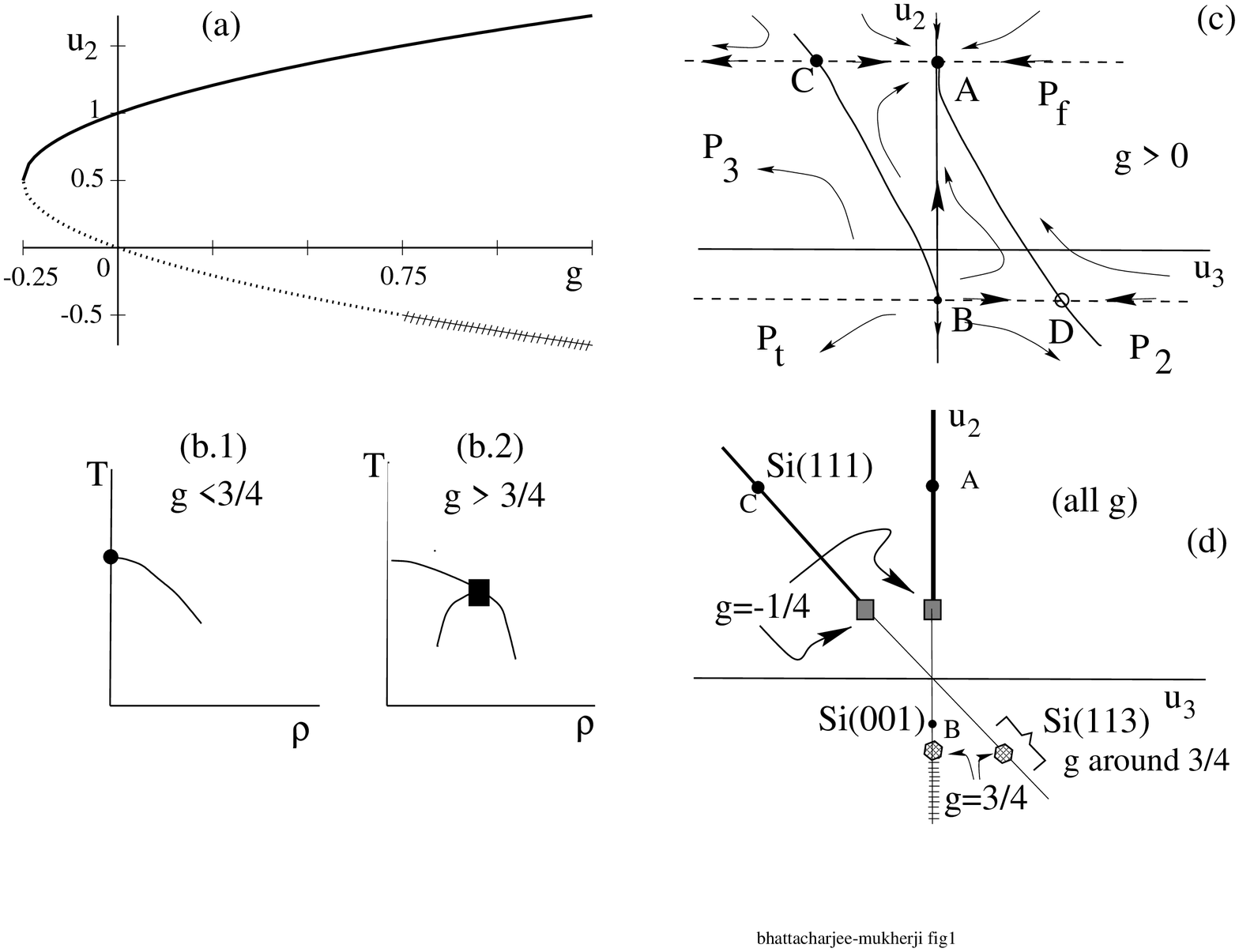,width=5in,angle=0}
\begin{figure}
 \caption{(a) Renormalization-group fixed-lines for $v_3=0$. These are 
   stable (thick line), unstable-critical (dotted), and
   unstable-first-order (hatched).  (b) A schematic orientational
   phase diagram in the $T$ vs $\rho$ plane for $u_2 \leq 0$.  For $g
   \leq 3/4$, there is a tricritical point (b.1) with $\beta\approx
   .5$.  For $g> 3/4$, a low-density first-order line connects the
   $\rho = 0$ transition to the coexistence curve either at the $\rho
   \neq 0$ critical-point of the latter or at an off-critical point
   (b.2).  (c) The fixed points in the $(u_2,u_3)$ plane for a $g >
   0$. Arrows indicate the RG flows.  The phases are $P_f$
   (uniformly-stepped, fractional-statistics like), $P_3$ (a phase
   with 3-steps bunching but no pairs), $P_2$ (pairs of steps but no
   triplets), and $P_t$ (all possible bunches).  (d) The fixed lines
   on a $u_2-u_3$ plane as $g(\geq -1/4)$ is varied.  Si surfaces
   showing similar behavior are indicated (not necessarily having same
   $g$).  Different types of lines represent different behaviors as
   noted in (a) and (c).
   ( Fig. 3 shows the fixed lines in a three-dimensional $(g,u_2,u_3)$ 
   plot.)}
 \label{fig:1}
\end{figure}
\begin{center}
\psfig{file=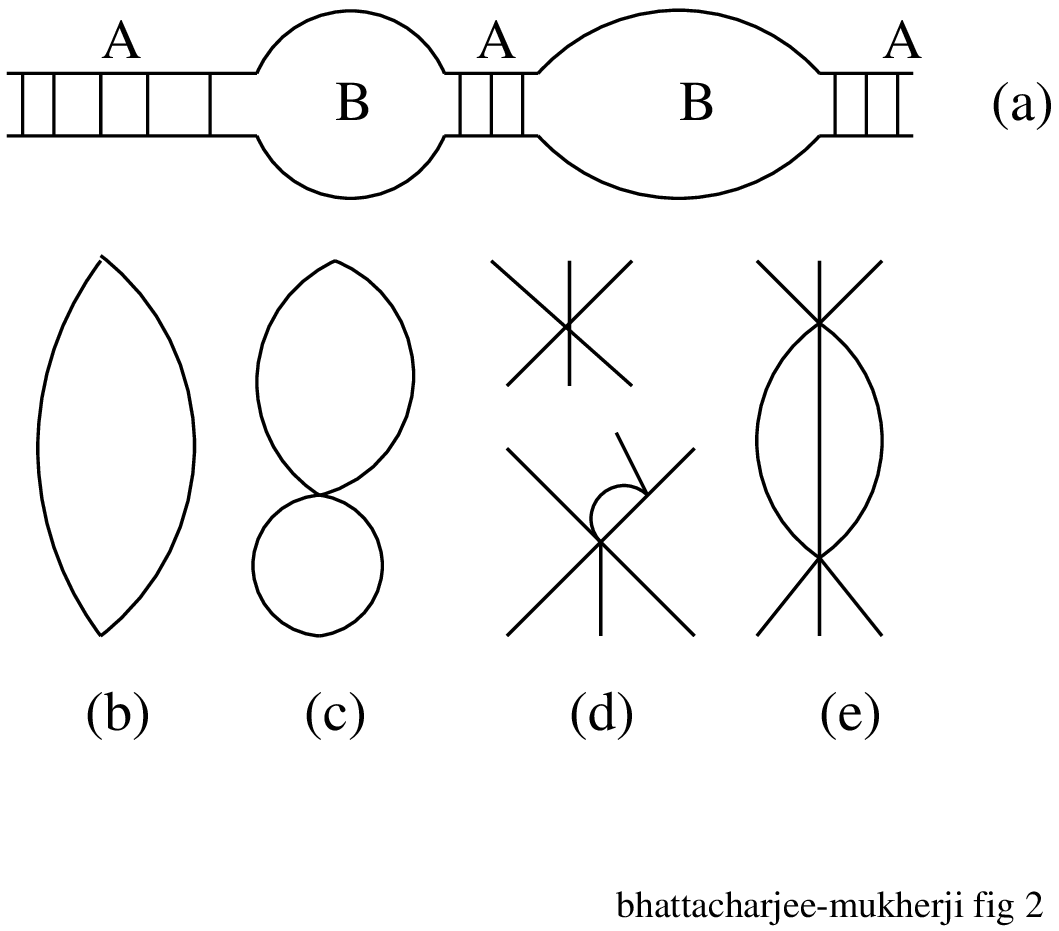,width=4in,angle=0}
 \end{center}
\begin{figure}
\caption{    (a) \ A typical configuration of two steps consisting of
  sequences of bound (A) and unbound (B) fragments of various lengths.
  (b, c) A zeroth-order and a first-order reunion diagrams of two
  steps.  (d) and (e) show the first- and second-order diagrams for
  $v_3$ renormlization.}
    \label{fig:2}
\end{figure}
}
\newpage
\vbox{
  \begin{figure}[htbp]
    \begin{center}
      \psfig{file=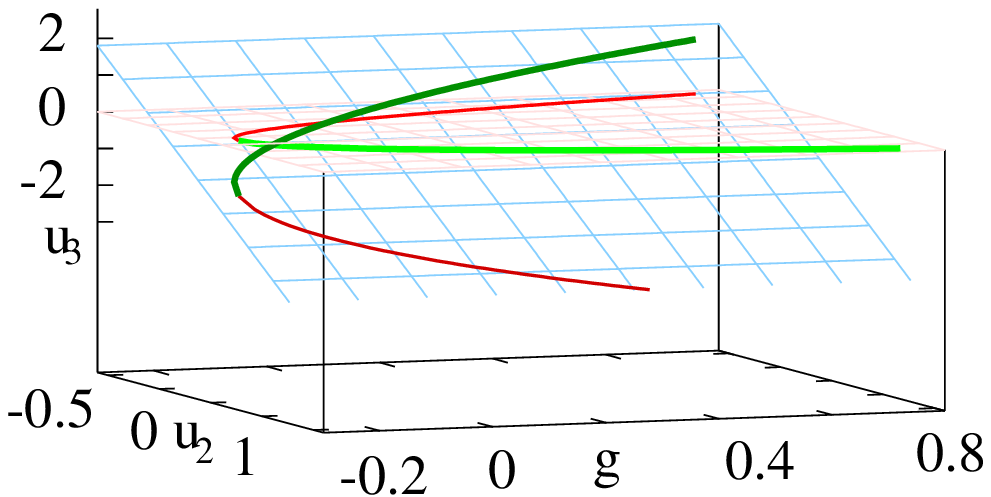,width=5in,angle=0} 
       \caption{The fixe lines in a $(g,u_2,u_3)$ space. Pairs are 
                 shown by similar colors.}  \label{fig:3}
    \end{center} 
\end{figure} 
}


\begin{references}
\bibitem{comm2} There are log corrections at $\psi =2$ 
  (or any higher integer).
\end{references}
 \end{document}